\documentstyle[preprint,aps]{revtex}

\begin{document}

\title{New constraints on ultrashort-ranged Yukawa interactions \\ from atomic force
microscopy\footnote{To appear in {\em Physical Review D}.}}

\author{
E.~Fischbach${}^{1,}$\footnote{Electronic address: ephraim@physics.purdue.edu}, 
D.~E.~Krause${}^{2,}$\footnote{Electronic address: kraused@wabash.edu},
V.~M.~Mostepanenko${}^{3,}$\footnote{On leave from A.Friedmann Laboratory
for Theoretical  Physics, St.Petersburg, Russia. 
Electronic  address:  mostep@fisica.ufpb.br}, and
M.~Novello${}^{3,}$\footnote{Electronic  address: novello@cbpf.br}
}
\address
{
${}^1$Physics Department, Purdue University, West Lafayette,
Indiana 47907--1396 \\
${}^2$Physics Department, Wabash College, Crawfordsville,
Indiana 47933--0352 \\
${}^3$Centro Brasileiro de Pesquisas F\'{\i}sicas, Rua Dr.~Xavier Sigaud, 150 \\
Urca 22290--180, Rio de Janeiro, RJ --- Brazil
}

\date{\today}

\maketitle

\begin{abstract}
Models in which the gravitational and gauge interactions are unified
 at $\sim 1$~TeV lead to the possibility that
large extra dimensions would produce
Yukawa-type corrections to the Newtonian gravitational law at small
distances.  In some models with $n = 3$ extra dimensions, deviations from
Newtonian gravity would occur at separations
$\sim~5$~nm, a distance scale accessible to an atomic force microscope
(AFM). Here we present constraints on
the Yukawa corrections derived from the latest AFM Casimir force
measurement by Mohideen, {\em et al.}
which are up to 19 times stronger than those obtained from their previous
experiment.   We then discuss new
designs for AFM experiments which have the potential to significantly
improve upon these constraints.
\end{abstract}



\section{Introduction}

Motivated in part by the hierarchy problem, a number of theoretical
models have been developed recently which raise the possibility that the
energy scale at
which gravity is unified with the other fundamental forces could be as low
as $\sim 1$ TeV
\cite{ADD,RS,extra dimension review}.  In these models, large extra spatial
dimensions
(which for compact dimensions could be as large as $\sim 1$ mm) arise which might be
accessible experimentally
\cite{Nima}.  While the effects of these new dimensions are generally model
dependent, constraints on extra
dimensional physics  which are relatively independent of theory can be
obtained from experiments searching
for deviations from Newton's law  of gravity at sub-millimeter distances.
Such experiments can be
classified according to the dominant background force acting between the
test bodies used.  For separations
$\gtrsim10^{-4}\,$m gravity produces the dominant background and a number of
groups are conducting experiments testing Newtonian gravity at $\sim 0.1$
mm scales.  For smaller
separations, Casimir forces provide the dominant background force, and the
experiments setting the best
limits in this regime are those testing the Casimir force law. In this
paper, we will briefly review
the theoretical motivation and phenomenology used in sub-millimeter force
experiments.
We then present new limits
extracted from a recent experiment by the Riverside group \cite{Mohideen 2000}, which
uses an atomic force microscope
to  test the Casimir force.   After comparing these results with
existing limits, we  conclude by describing future
possibilities for using AFM experiments to set limits on new forces and extra dimensional
physics.

\section{Phenomenology}

Since gravity is a theory of spacetime, it is inevitable that the effects
of extra dimensions
will modify Newton's law of gravity which is only valid in a 4-dimensional
spacetime.
Although Newtonian  gravity has been well-tested over long distance scales,
our understanding of gravity at
sub-millimeter scales is quite limited \cite{Fischbach book}.
Phenomenologically, the
deviations from Newton's law arise naturally in two different ways
\cite{Chung}.  In models with large
compact dimensions such as originally
proposed by Arkani-Hamed, Dimopoulos, and Dvali (ADD) \cite{ADD}, one finds
that the potential energy between two point particles with masses $m_{1}$ and
$m_{2}$ separated by a
distance $r \gg R_{*}$ (the characteristic size of the compact dimensions) is given by the
usual Newtonian gravitational potential energy with a Yukawa correction
\cite{Nima,Kehagias,Floratos}:
\begin{equation}
V(r) = -\frac{Gm_{1}m_{2}}{r}\left(1 + \alpha e^{-r/\lambda}\right);
\phantom{space}(r \gg R_{*}),
\label{Yukawa}
\end{equation}
where the dimensionless constant $\alpha$
depends on the nature of the extra dimensions, and where $\lambda \sim R_{*}$.  For
example, for a toroidal compactification with all
$n$ extra dimensions having equal size, $\alpha = 2n$
\cite{Nima,Kehagias,Floratos}.   When the separation between the masses decreases to the
point where $r \ll R_{*}$, the usual inverse square law of gravity changes to a new
power-law:
\begin{equation}
V(r) = -\frac{G_{4+n}m_{1}m_{2}}{r^{n + 1}}
\phantom{space}(r \ll R_{*}),
\label{ADD power-law}
\end{equation}
where $G_{4+n} \sim GR_{*}^{n}$ is the fundamental gravitational
constant  in the full $4 + n$ dimensional spacetime.
The size $R_{*}$ of the extra dimensions, according to the ADD models, is then related to
the energy scale $M_{*} \sim (1/G_{4+n})^{1/(2 +n)}$ at which the unification occurs by
\begin{equation}
R_{*} \sim \frac{1}{M_{*}}\left(\frac{M_{Pl}}{M_{*}}\right)^{2/n} \sim
10^{32/n - 17}\,\,\mbox{cm},
\label{R*}
\end{equation}
where the Planck mass is $M_{Pl} = 1/G^{1/2} \sim 10^{19}$ GeV, and we have assumed
$M_{*} \sim 1$~TeV~\cite{ADD}.  For
$n =1$, $R_{*}
\sim 10^{15}$ cm is obviously excluded by solar system tests of Newtonian
gravity \cite{Fischbach book}.
If $n = 2$, $R_{*} \sim 1$ mm, which is the scale currently being probed by a number of
sub-millimeter gravity
experiments \cite{Hoyle,Physics Today,Long,Carugno}.
For $n = 3$, Eq.~(\ref{R*})  gives $R_{*} \sim 5$ nm which is  roughly the
range accessible to experiments
using atomic force microscopy like those  discussed  below.    It
is important to recognize,
however, that while $M_{*} \sim 1$ TeV is a natural scale,  it is 
little more than a reasoned
guess.  Therefore, the scales suggested by Eq.~(\ref{R*}) should be taken
simply as a heuristic guide for designing
experiments.  It should also be noted that the Yukawa correction given in
Eq.~(\ref{Yukawa}) arises naturally
in ways unrelated to extra dimensional
physics.  For example, Yukawa
potentials arise from new forces generated by the exchange of bosons of mass
$\mu = 1/\lambda$ (e.g., scalar, graviphoton, dilaton)
\cite{Fischbach book,Fujii,Adelberger Review,Giudice}.

	A second class of extra dimensional models, such as that originally
proposed by Randall and Sundrum
\cite{RS}, allow the extra dimensions to be non-compact, but warped.   In
these models, the leading
order correction to the Newtonian potential energy between two point masses
takes  on a power-law
form \cite{RS,Chung,Garriga}:
\begin{equation}
V(r) = -\frac{Gm_{1}m_{2}}{r}\left(1 + \frac{2}{3k^{2}r^{2}}\right),
\phantom{space}(r \gg 1/k),
\label{RS power-law}
\end{equation}
where $1/k$ is the warping scale.  The introduction of a new length scale
$r_{0}$ such as
$1/k$ or $R_{*}$ into a theory allows one to generalize Eq.~(\ref{RS
power-law}) to arbitrary
powers $N$:
\begin{equation}
V(r) = -\frac{Gm_{1}m_{2}}{r}\left[1 +
\alpha_{N}\left(\frac{r_{0}}{r}\right)^{N-1}\right],
\label{general power-law}
\end{equation}
where $\alpha_{N}$ is a dimensionless constant.   Power-law corrections to
Newtonian gravity
also arise from new interactions involving the exchange of 2 massless
quanta.  For example,
$N = 2$ may arise from the simultaneous exchange of two photons or two
massless
scalars \cite{FS power-law}, and $N = 3$ characterizes the
exchange of two massless pseudoscalars \cite{pseudoscalars,Grifols}.
Potentials with $N = 5$ arise from massless axion exchange \cite{Grifols},
and also from the
exchange of a (massless) neutrino-antineutrino ($\nu\overline{\nu}$) pair
\cite{FS neutrino}.

	All of the extra dimensional models discussed above are speculative
at this
stage, and much work remains to be done to test their consistency and
viability.  However, they are important in demonstrating that if extra spatial
dimensions
exist, new fundamental energy scales must also exist, and that these
scales will modify
Newtonian gravity at some level.  Furthermore, from the models that have
been developed we
see that these modifications generally take the form of Yukawa or power-law
potentials.
Therefore, while there remain many unanswered questions regarding models of
extra-dimensional
physics, the phenomenology supporting short distance gravity experiments
rests upon a much
firmer footing.

As noted earlier, experiments testing Newtonian gravity over
separations $\gtrsim~10^{-3}\,$m have set significant limits on the Yukawa
coupling $\alpha$ for
ranges $\lambda \gtrsim 10^{-3}$ m \cite{Fischbach book,Smith}.  In
addition, precise gravity
experiments have set the best limits on power-law interactions for $N < 5$
(see Table
\ref{power-law limit table}).
Motivated in part by the new extra dimensional models discussed above,
several experimental
groups \cite{Hoyle,Physics Today,Long,Carugno} are now probing  Newtonian
gravity at
sub-millimeter scales and have begun to set significant constraints on new
physics at ranges
$\sim 0.1$ mm.  It is expected that within the next few years such
experiments will extend
our understanding of gravity down to separations where Casimir forces form
the significant
background.

\section{New Forces and Tests of the Casimir Force}

Coincidentally, the past few years have also witnessed the first precise tests of
the Casimir force
\cite{Casimir book,Casimir bibliography,Milonni}.  Lamoreaux
\cite{Lamoreaux} used a torsion
balance to measure the Casimir force between a disk and spherical lens,
while  Ederth \cite{Ederth} measured the force between two crossed
cylinders.  The practical implications of the Casimir force have been considered 
recently with a MEMS (microelectromechanical) device by Chan, {\em et al.} at Bell Labs
\cite{Chan}.   Finally, a series of force measurements specifically
devised to test various aspects of  the Casimir force has been conducted by the Riverside
group using an atomic force microscope
\cite{Mohideen 2000,Mohideen 1998,Mohideen 1999a,Mohideen 1999b}.   This entire
experimental effort has motivated
theorists to more carefully examine the effects of surface roughness,
temperature, and finite
conductivity which produce significant deviations from the idealized case
of the Casimir force between two perfectly conducting, smooth bodies
\cite{Klim 01,Klim 60,Klim 61,Bezerra,Bordag PRL,Genet,Lambrecht}.  The inclusion
of these corrections has
been shown to be crucial in obtaining agreement between theory and experiment
\cite{Mohideen 2000,Mohideen
1998,Mohideen 1999a,Mohideen 1999b,Klim 60,Klim
61,Lamoreaux Comment,Mohideen and Roy Comment}.

While the goal of all these experiments is to improve our understanding of the
origin and behavior of  Casimir forces, they have
also been used to set limits on new interactions \cite{Bordag 1998,Bordag
1999,Bordag 2000,Most and Novello 2001}.  These limits are in addition to those
obtained in Refs.~\cite{Most 1988,Most 1993,Bordag 1994}
from earlier, much
less precise Casimir and van der Waals force measurements \cite{Israel,Derjaguin}.
It can be shown that Casimir force measurements are relatively ineffective compared to
longer ranged gravity
experiments in setting limits on power-law interactions of the form given
by Eq.~(\ref{general
power-law}).  The reason is that for $N < 5$, such power-law forces  between
macroscopic bodies are
relatively insensitive to the separation of the bodies  when these are
placed in close proximity \cite{Bordag
1999}.  Thus, there is no advantage to performing very short distance
experiments, which are in any case
generally less precise than larger scale gravity experiments, to search for power-law
corrections to Newtonian gravity.

On the other hand, experiments searching for  Yukawa interactions of the
form given by
Eq.~(\ref{Yukawa}) are only sensitive when the separation $r$ of the test
bodies is of order  $\lambda$
\cite{Fischbach book}.  Hence, to constrain the Yukawa coupling $\alpha$
for a given value of
$\lambda$, one must devise experiments such that $r$ is of order $\lambda$.
Therefore, if one wishes to
constrain extra dimensions of size $R_{*} \lesssim 10^{-5}$~m using force
measurements, one must
inevitably confront large Casimir forces which  grow rapidly for $r
\lesssim 10^{-5}$~m \cite{Krause}.  This is why Casimir force experiments
yield the best
constraints on $\alpha$ for
these separations.

\section{Limits from Recent Mohideen AFM Experiment}

We turn next to the focus of this paper, which is the recent measurement of
the Casimir force
performed by the Riverside group \cite{Mohideen 2000}  who used an atomic force
microscope. (We will henceforth refer to the work of the Riverside group as ``Mohideen,
{\em et al.}'') Like the previous AFM measurements carried out by this group \cite{Mohideen
1998,Mohideen 1999a,Mohideen 1999b}, this experiment measured the force
between a
metallized polystyrene sphere mounted on an AFM tip and a substrate composed
of a flat sapphire
disk.  However, this experiment differed from the previous efforts in
several important respects:
(1) The sphere and  disk were each coated with a layer of gold of thickness
$\Delta = 86.6$ nm.  Previously, aluminum coatings had been used instead, but this
also required an
additional  surface layer of Au/Pd on top of the Al  to prevent oxidation.
The use of only a
single gold layer in the new experiment significantly simplifies
the calculations of the forces involved, since one can treat
the gold layer as infinitely thick as far as the Casimir force is
concerned.  Also, because Au is much more
dense than the previously
used layers, stronger limits on new forces can be obtained.
(2) The gold layers were significantly smoother than previous coatings.
The 
root-mean-square amplitude of the gold surfaces  was measured to be only
$1.0 \pm 0.1$ nm which
 means that corrections arising from the surface roughness could be
neglected in all force
calculations. (3) Electrostatic forces which plague nearly all short
distance force measurements
were reduced to $\ll 1$\% of the Casimir force at the shortest separation.
This meant that no
subtractions were required to separate Casimir forces from spurious
electrostatic background
forces. The electrostatic force was used to arrive at an independent
measurement of the surface separation including separation on contact
of the two surfaces.  
(4) The  measurements were performed over smaller separations $a$,
$62\,\mbox{nm}\leq a \leq 350$~nm which means this experiment can search
for Yukawa forces with
smaller ranges $\lambda$.  However, the absolute error of the force
measurements was somewhat larger
than in the previous experiments, $\Delta F = 3.5\times 10^{-12}\,$N.  This is due to
the thinner gold coating used in \cite{Mohideen 2000} which led to poor thermal
conductivity of the
cantilever.  At the smallest measured separations, this error was still
less than 1\% of the
measured Casimir force.  Thus, this experiment can be used to significantly
tighten constraints
on new Yukawa interactions as we will now show.

 We begin by calculating the force arising between the AFM tip and the
substrate using
Eq.~(\ref{Yukawa}).  First, it is easy to show that the Newtonian
contribution is negligible.  Since the diameter of the sphere (including
the gold layer) was
$2R = 191.3$ $\mu$m, while the disk diameter was $L = 1$ cm, it follows that $R \ll L$
which allows us to consider each atom of the sphere as if it were placed directly above the
center of the disk.
In this case, the vertical
component of the Newtonian gravitational force acting between the disk and
a sphere
atom of mass
$m_1$ located at a distance $l\ll L$ above the disk is
\begin{eqnarray}
f_{N,z}(l) & = & \frac{\partial}{\partial l}\left[
Gm_1\rho_{\rm disk}2\pi
\int\limits_{0}^{L}r\,dr
\int\limits_{l}^{l+D}\frac{dz}{\sqrt{r^2+z^2}}\right]
\nonumber \\
& \approx&  -2\pi Gm_1\rho_{\rm disk} D\left[
1-\frac{D+2l}{2L}\right],
\label{11}
\end{eqnarray}
where $\rho_{\rm disk}$ is the disk density, $D = 1$ mm is the thickness of the disk, and
only the first order terms in
$D/L$ and $l/L$ have been retained. The Newtonian gravitational force
acting between the disk and
the sphere is obtained from Eq.~(\ref{11}) by integration over the sphere
\begin{equation}
F_{N,z}\approx -\frac{8}{3}\pi^2G\rho_{\rm disk}\rho_{\rm sphere}DR^3\left(
1-\frac{D}{2L}-\frac{R}{L}\right),
\label{12}
\end{equation}
where $\rho_{\rm sphere}$ is the density of the sphere.
Even if the sphere and disk were composed of solid vacuo-distilled gold
with $\rho_{\rm disk}=\rho_{\rm sphere}=18.88\times10^3\,$kg/m${}^3$, one
finds from
Eq.~(\ref{12}) that $F_{N,z}\approx 6\times 10^{-16}\,\mbox{N}\ll\Delta F$,
so the
usual Newtonian gravitational force can be neglected.

The force arising from the Yukawa correction to  Newtonian gravity given by
the
second term of Eq.~(\ref{Yukawa}) should be calculated taking into account
the actual compositions of the test bodies.   For a polystyrene sphere
$\rho_{\rm sphere}=1.06\times10^3\,$kg/m${}^3$,
for a sapphire disk $\rho_{\rm disk}=4.0\times10^3\,$kg/m${}^3$,
and for the vacuo-distilled gold covering layers
$\rho_{\rm Au}=18.88\times10^3\,$kg/m${}^3$.
The Yukawa force $F_{Y}$ arising from Eq.~(\ref{Yukawa}) can be  easily obtained
using the same procedure
used to calculate the Newtonian gravitational force. The result is
\begin{eqnarray}
&&F_{Y}(a)=-4\pi^2G\alpha\lambda^3e^{-a/\lambda}R
\nonumber \\
&&\phantom{aaa}\times\left[\rho_{\rm Au}-(\rho_{\rm Au}-\rho_{\rm
disk})e^{-\Delta/\lambda}
\right]\, \left[\rho_{\rm Au}-(\rho_{\rm Au}-\rho_{\rm
sphere})e^{-\Delta/\lambda}
\right],
\label{13}
\end{eqnarray}
where $a$ is the sphere-disk separation distance, and $\Delta = 86.6$ nm is the thickness
of the gold coatings.

The most stringent limits on new forces are obtained for the smallest possible separation.
According to Ref.~\cite{Mohideen 2000}, the rms deviation of the measured force ($F_{\rm
experiment}$) from the theoretical value ($F_{\rm theory}$) of the sum of all the known
forces (Casimir and electrostatic)  is given by
\begin{equation}
\sigma_{F} = \sqrt{\frac{\sum\left(F_{\rm experiment} - F_{\rm theory}\right)^{2}}{N_{\rm
trials}}} = \mbox{3.8 pN},
\label{F rms}
\end{equation}
where $N_{\rm trials} = 2583$ is the total number of force measurements.  This is
slightly larger than 3.5 pN,  the experimental uncertainty of the force measurement at
$a = 62$ nm.    Therefore, to set constraints on the Yukawa coupling constant
$\alpha$, we will assume
$|F_{Y}(a)| \leq  \sigma_{F} = $ 3.8 pN. The resulting limit is represented by the bold
curve labeled ``Mohideen 2000'' in Fig.~\ref{limit figure}.

Before comparing the constraint obtained here with limits obtained from other experiments,
let us first discuss various sources of error which could affect our result.  First, the
uncertainty in the separation $a$ arising from the surface roughness is approximately
$\delta a \simeq 1$ nm which gives $\delta a/a
\simeq 2\%$.  This will produce a significant effect when $\lambda \simeq \delta a \simeq
1$~nm, but as we will see below, other experiments produce better limits in this range. 
For $\lambda \simeq a$, where the most stringent limits are obtained,
the resulting errors are only $\sim 2\%$ which can be neglected.  Second, it is important
that all known forces be included in $F_{\rm theory}$ in using Eq.~(\ref{F rms}).  In short
distance force experiments, electrostatic effects can be significant.  However, in the
Mohideen experiment, the measured residual potential difference of  3 mV between the gold
sphere and disk leads to a force only 0.1\% of the Casimir force at $a = 62$ nm. 
Although this effect is small, it was included in
$F_{\rm theory}$.

	Finally, the most significant difficulties arising in extracting limits from Casimir force
experiments involve calculating the Casimir force between the interacting bodies. 
Due to practical difficulties, none of the most recent tests of the Casimir force use
parallel plates for the interacting bodies, for which accurate calculations can be made. 
(The experiment by Carugno \cite{Carugno} uses parallel plates, but has not yet reached
sufficient sensitivity to detect the Casimir force.)  Rather, curved surfaces are used
which requires the use of the proximity force theorem (PFT) to calculate the Casimir force
\cite{Blocki}.  Since the PFT produces reasonable results when applied to Casimir
forces, it is probably correct.   Although there has not
been a rigorous proof of the PFT in this context
\cite{Lambrecht} (see Ref.~\cite{Schaden} for an interesting semi-classical argument), 
we will assume that its use in the analysis of the Mohideen experiment is valid. 

 Other difficulties 
in Casimir force calculations arise from the use of tabulated values of the
complex dielectric constant as functions of frequency for the materials comprising the test
bodies.  It has been pointed out that if different interpolation schemes are applied to
these discrete data, the resulting calculated Casimir force can differ by as much as 4\%
\cite{Lambrecht,Bostrom}.   In addition, the tabulated dielectric properties were obtained
using bulk samples and not the coatings which are actually used in the experiment.  It
would then be best to directly measure the dielectric properties of the actual test
bodies to accurately determine $F_{\rm Casimir}$ \cite{Lamoreaux 1999}.  Since this cannot
be done in the present context,  and since there is no obvious discrepancy or
inconsistency arising in the Mohideen experiment, we will not consider this problem here
and leave it to future experiments to address this issue.

\section{Comparison with Other Limits}

      As can be seen from Fig.~\ref{limit figure}, the Casimir force measurement
between the gold surfaces carried out by Mohideen, {\em et al.}  strengthens the
previously known constraints obtained from their earlier experiment \cite{Mohideen 1999a}
using aluminum surfaces (labeled by ``Mohideen 1999'') by as much as a factor of 19 
within the range
$4.3\times 10^{-9}\,\mbox{m}\leq\lambda\leq 1.5\times 10^{-7}\,$m, with the most
significant improvement occurring  at $\lambda=$(5--10)\,nm.
These constraints are up to 4500 times more stringent than those obtained
from older Casimir and van der
Waals force measurements between dielectrics (curves labeled by ``Casimir'' and ``vdW''
respectively). In the same figure, the curve labeled by ``Lamoreaux'' exhibits the
constraint  obtained in Ref.~\cite{Bordag 1998} from the
Casimir force measurement
which used a torsion pendulum \cite{Lamoreaux}.  
The curve denoted by ``Ederth'' gives the new constraints obtained in Ref.~\cite{Most and
Novello 2001} from the recent experiment which measured the Casimir force between two
crossed cylinders
\cite{Ederth}. We see that curve ``Mohideen 2000,'' which exhibits the limits obtained in
this paper, represents the most stringent bound in the interaction range $1.1\times
10^{-8}\,\mbox{m}\leq\lambda\leq 1.5\times 10^{-7}\,$m.

It is evident from Fig.~\ref{limit figure} that for separations $\lesssim 10^{-5}$ m
much work is needed to improve the sensitivity of force experiments to Yukawa interactions
of the strength predicted by extra dimensional models ($\alpha \sim 1$--10).   As was shown in Ref.~\cite{Bordag
1998}, the constraints following from the experiment in Ref.~\cite{Lamoreaux} can be
improved by up to four orders of magnitude in the range
$\lambda \sim 10^{-4}$~m which is what is required to
constrain physically interesting values of $\alpha$. Still, experiments using 
atomic force microscopy \cite{Mohideen 2000,Mohideen 1998,Mohideen 1999a,Mohideen
1999b} remain almost fifteen orders of magnitude below the sensitivity
needed to reach the value $\alpha \sim$1--10 in the interaction range
$\lambda\leq 10^{-7}\,$m.  Furthermore, significant (and sometimes even more stringent)
constraints on the extra dimensional mass scale $M_{*}$ have been obtained
from accelerator experiments \cite{Abbott,LEP,HERA}, astrophysics
\cite{Hanhart,Barger,Cullen,Hannestad,Cassisi}, and cosmology
\cite{Hall,Fairbairn,Hannestad2}.  However, one
must remember that gravity and Casimir force experiments are also sensitive to other
non-extra dimensional effects such as dilaton and moduli exchange which can lead to
Yukawa forces with $\alpha \gg 10$ \cite{Fischbach book,Fujii,Adelberger Review,Giudice}. 
Therefore, all experiments which can reduce the allowed region in the $\alpha$--$\lambda$
plane are meaningful.

\section{Future AFM Experiments}

We believe that in the future the best ultrashort range limits
will come from a new class of
experiments designed specifically to search for new forces rather than to 
test the Casimir force.  In
such experiments the Casimir effect is an unwanted background  that
needs to be suppressed.  One
suppression technique proposed recently \cite{Krause} relies on the observation
that the Casimir force depends on the {\em electronic} properties of
interacting samples, whereas any
gravitational interactions (either conventional or new), and
virtually all other proposed interactions, involve couplings to
nuclei (as well as to electrons).  Hence by comparing the
interactions of two test masses which have very similar electronic
properties, but different nuclei, with a common attracting source
mass, we may be able to subtract out the common Casimir background.
This would leave a residual nuclear-nuclear interaction between each
of the samples and the source, and these should be different for the
two samples.  We refer to this technique as the {\em iso-electronic
effect}, and it can be implemented in two ways.  One is to choose as
the test masses elements such as Cu and Au whose nuclei are quite
different, but which are known to have very similar Casimir properties
\cite{Lambrecht}.  Alternatively, the test samples could be chosen to
be isotopes of the same element whose nuclei would contain different
numbers of neutrons, but whose electronic properties would be
extremely similar.  In either case, any residual differences in the
electronic properties of the test masses would still have to be
calculated, but this should in principle present no significant
problems \cite{Lambrecht personal communication}.

A possible design for an AFM experiment \cite{STEP paper,Howell} which utilizes the
iso-electronic effect, is shown
schematically in Fig.~\ref{design figure}.  In this
experiment, a gold-coated sphere would be attached to an AFM tip and
oscillated with frequency $\omega_{\rm tip}$.  The substrate would be
composed of alternating strips of different isotopes of the same
element, and would be oscillated horizontally with frequency
$\omega_{\rm sub}$.  The signal for a new force would then be a force
on the tip that depends on $\omega_{\rm sub}$ which should not be
present if surface roughness and electrostatic effects are the same
for the strips.  Since the Yukawa force of the AFM tip on the substrate
couples only to the mass within
a range $\lambda$ of the surfaces, the net sphere-substrate force will be
proportional to the difference in
the  densities of the strips:
\begin{equation}
F_{\rm Yukawa}  \propto  (\rho_{\rm sub} - \rho_{\rm sub}').
\end{equation}
Therefore, it is important to select elements having isotopes with the
largest possible mass density
differences.  Possible candidates include ruthenium, osmium, nickel, and
palladium, all of which can have
isotope mass density differences exceeding $900$ kg/m$^{3}$. Other possible designs
for ultrashort distance experiments can be found in Refs.~\cite{Long,Krause,STEP paper}. 

\section{Conclusions}

It is clear that experiments which search for new forces in the Casimir
regime will be faced with
significant challenges.  However, there is strong theoretical
motivation to conduct such searches, and new techniques (including, possibly, those
suggested here) will certainly be developed to overcome these
obstacles.  The AFM is a natural tool to be used in this effort as demonstrated by the
experiments of Mohideen, {\em et al.} \cite{Mohideen 2000,Mohideen 1998,Mohideen
1999a,Mohideen 1999b}.  Furthermore, this instrument can be easily
adapted to experiments whose intent is to specifically search for new forces.
Such laboratory force experiments have provided, and will continue to
provide, important, and relatively
model-independent  constraints on new physics which complement those
obtained from high-energy experiments
and astrophysical observations.

\acknowledgements
The authors thank A. Lambrecht, S. Howell, G.~L.~Klimchitskaya,
U.~Mohideen, R. Reifenberger, and S. Reynaud for helpful discussions.
We also thank S. Karunatillake and M. West for their calculations of the isotopic mass
densities.  V.~M.~M. is grateful to the
Centro Brasileiro de Pesquisas F\'{\i}sicas for their kind hospitality.
This work was supported
in part by the U. S. Department of Energy under contract No.
DE-AC02-76ER071428 (E. F.), the
Wabash College Byron K. Trippet Research Fund (D. E. K.), FAPERJ
 and CNPq
(V.~M.~M. and M.~N.).

\newpage

\begin{table}
\begin{center}
\caption{Current constraints on power-law potentials of the form given by
Eq.~(\ref{general power-law}).}
\begin{tabular}{cccc}
$N$ & Experiment & $\alpha_{N}r_{0}^{N-1}$ & Limit Reference \\ \hline
1 & Gundlach, {\em et al.}\cite{Gundlach} &  $1 \times 10^{-9}$ &
\cite{Gundlach}\\
2 & Gundlach, {\em et al.}\cite{Gundlach} &  $7 \times 10^{-7}$ m$^{1}$  &
\cite{Gundlach}\\
3 & Spero, {\em et al.}\cite{Hoskins} & $1 \times 10^{-8}$ m$^{2}$ &
\cite{Fischbach and Krause}\\
4 & Mitrofanov, {\em et al.}\cite{MP} & $1 \times 10^{-10}$ m$^{3}$
&\cite{Fischbach and Krause}
\end{tabular}
\label{power-law limit table}
\end{center}
\end{table}

\newpage

\begin{figure}
\caption[]{Constraints on the Yukawa interactions of the form given by Eq.~(\ref{Yukawa})
from Casimir/van der Waals force measurements.  ``Mohideen 2000'' refers to the limit
obtained in this paper from the latest AFM experiment by the Riverside group
\cite{Mohideen 2000}, while ``Mohideen 1999'' refers to the limit \cite{Bordag
2000} obtained 
 from their previous AFM experiment  which used aluminum surfaces \cite{Mohideen 1999a}. 
``vdW'' is the limit \cite{Bordag 1994} obtained from older van der Waals force experiments
\cite{Israel}, ``Ederth'' is the limit \cite{Most and Novello 2001} from the recent
experiment by Ederth
\cite{Ederth} using two crossed cylinders, ``Casimir'' is the limit \cite{Most 1988} from
older Casimir force experiments \cite{Derjaguin}, and ``Lamoreaux'' is the limit
\cite{Bordag 1998} from the Casimir force experiment by Lamoreaux \cite{Lamoreaux}. }
\label{limit figure}
\end{figure}

\begin{figure}
\caption[]{Proposed experiment  utilizing the iso-electronic effect \cite{STEP
paper,Howell}. A gold-coated sphere is attached to an oscillating AFM tip, while a
substrate composed of alternating strips of different isotopes of the
same element is oscillated horizontally beneath.}
\label{design figure}
\end{figure}



\begin{thebibliography}{99}


\bibitem{ADD} N. Arkani-Hamed, S. Dimopoulos, and G. Dvali, Phys. Lett. B
{\bf 429},
263 (1998).


\bibitem{RS}  L. Randall and R. Sundrum, Phys. Rev. Lett. {\bf 83}, 3370
(1999); L.
Randall and R. Sundrum,  Phys. Rev. Lett. {\bf 83}, 4690 (1999).

\bibitem{extra dimension review} A. P\'{e}rez-Lorenzana, hep-ph/0008333.


\bibitem{Nima} H. Arkani-Hamed, S. Dimopoulos, and G. Dvali, Phys. Rev. D
{\bf 59}
086004 (1999).

\bibitem{Mohideen 2000} B. W. Harris, F. Chen, and U. Mohideen, Phys. Rev.
A {\bf 62},
052109 (2000).


\bibitem{Fischbach book} E.~Fischbach and C.~L.~Talmadge,
{\it The Search for Non-Newtonian Gravity} (Springer-Verlag, New York, 1999).



\bibitem{Chung}  D. J. H. Chung, L. Everett, and H. Davoudiasl, hep-ph/0010103.

\bibitem{Kehagias} A. Kehagias and K. Sfetsos, Phys. Lett. B {\bf 472}, 39
(2000).

\bibitem{Floratos}  E. G. Floratos and G. K. Leontaris, Phys. Lett. B {\bf
465}, 95
(1999).


\bibitem{Hoyle} C. D. Hoyle, U. Schmidt, B. R. Heckel, E. G. Adelberger, J.
H. Gundlach,
D. J. Kapner, and H. E. Swanson, Phys. Rev. Lett. {\bf 86} 1418 (2001).

\bibitem{Physics Today} B. Schwarzschild, Physics Today {\bf 53} (9), 22
(2000);
R. Newman, Matters of Gravity, No. 15, 18 (2000); R. Newman, Matters of
Gravity, No. 16, 12 (2000).

\bibitem{Long} J. C. Long, A. B. Churnside, and J. C. Price,
hep-ph/0009062; J.~C.~Long,
H.~W.~Chan, and J.~C.~Price, Nucl. Phys. B {\bf 539}, 23 (1999).

\bibitem{Carugno} G. Carugno, Z. Fontana, R. Onofrio, and C. Rizzo, Phys.
Rev. D {\bf
55}, 6591 (1997); G. Bressi, G. Carugno, A. Galvani, R. Onofrio, and G.
Ruoso, Class. Quantum
Grav. {\bf 17} (2000) 2365.


\bibitem{Fujii} Y. Fujii, Int. J. Mod. Phys. A {\bf 6}, 3505 (1991).

\bibitem{Adelberger Review} E. G. Adelberger, B. R. Heckel, C. W. Stubbs, and W. F.
Rogers, Annu. Rev. Nucl. Part. Sci. {\bf 41}, 269 (1991).

\bibitem{Giudice} S. Dimopoulos and G. F. Giudice, Phys. Lett. B {\bf 379}, 105 (1996).

\bibitem{Garriga} J. Barriga and T. Tanaka, Phys. Rev. Lett. {\bf 84}, 2778
(2000).

\bibitem{FS power-law} J. Sucher and G. Feinberg, in {\em Long-Range
Casimir Forces}, ed. by F. S.
Levin and D. A. Micha  (New York, Plenum, 1993).

\bibitem{pseudoscalars}  S. D. Drell and K. Huang, Phys. Rev. {\bf 91},
1527 (1953);
V. M. Mostepanenko and I. Yu. Sokolov Sov. J. Nucl. Phys. {\bf 46}, 685 (1987).

\bibitem{Grifols} F. Ferrer and J. A. Grifols, Phys. Rev. D {\bf
58}, 096006 (1998).

\bibitem{FS neutrino} G. Feinberg and J. Sucher, Phys. Rev.
{\bf 166} 1638 (1968); S. D. H. Hsu and P. Sikivie, Phys. Rev. D {\bf 49}
4951 (1994);
E. Fischbach, Ann. Phys. (NY) {\bf 247}, 213 (1996).

\bibitem{Smith} G.~L.~Smith, C.~D.~Hoyle, J.~H.~Gundlach, E.~G.~Adelberger,
B.~R.~Heckel, and H.~E.~Swanson, Phys. Rev. D {\bf 61}, 022001 (1999).

\bibitem{Gundlach}
J. H. Gundlach, G. L. Smith, E. G. Adelberger, B. R. Heckel, and
H. E. Swanson,
Phys. Rev. Lett. {\bf 78}, 2523 (1997).

\bibitem{Hoskins}
J. K. Hoskins, R. D. Newman, R. Spero, and J. Schultz,
Phys. Rev. D {\bf 32}, 3084 (1985).

\bibitem{MP}
V.~P.~Mitrofanov and O.~I.~Ponomareva, Sov. Phys. JETP (USA) {\bf 67}, 1963
(1988).

\bibitem{Fischbach and Krause} E. Fischbach and D. E. Krause, Phys. Rev. Lett. {\bf 83}, 3593 (1999).

\bibitem {Casimir book} V.~M.~Mostepanenko and N.~N.~Trunov, {\it The
Casimir Effect
and Its Applications} (Clarendon, Oxford, 1997).

\bibitem{Casimir bibliography} S. K. Lamoreaux, Am. J. Phys. {\bf 67} 850
(1999).

\bibitem{Milonni} P. W. Milonni, {\it The Quantum Vacuum} (Academic Press,
San Diego, 1994).

\bibitem{Lamoreaux} S. K. Lamoreaux, Phys. Rev. Lett. {\bf 78}, 5 (1997);
{\bf 81}, 5475(E) (1998).

\bibitem{Ederth} T. Ederth, {\em Phys. Rev. A} {\bf 62}, 062104 (2000).

\bibitem{Chan} H. B. Chan, V. A. Aksyuk, R. N. Kleiman, D. J. Bishop, and F. Capasso,
Science {\bf 291}, 1941 (2001).

\bibitem{Mohideen 1998} U. Mohideen and A. Roy, Phys. Rev. Lett. {\bf 81},
4549 (1998).

\bibitem{Mohideen 1999a} A. Roy, C. -Y. Lin, and U. Mohideen Phys. Rev. D
{\bf 60} 111101(R)
(1999).

\bibitem{Mohideen 1999b} A. Roy and U. Mohideen, Phys. Rev. Lett. {\bf 82}
4380 (1999).

\bibitem{Klim 01} G.~L.~Klimchitskaya and
V.\ M.\ Mostepanenko, Phys. Rev. A {\bf 63}, 0621XX (2001).


\bibitem{Klim 60} G.~L.~Klimchitskaya, A. Roy, U.~Mohideen, and
V.\ M.\ Mostepanenko, Phys. Rev. A {\bf 60}, 3487 (1999).

\bibitem{Klim 61} G.~L.~Klimchitskaya,  U.~Mohideen, and
V.\ M.\ Mostepanenko, Phys. Rev. A {\bf 61}, 062107 (2000).

\bibitem{Bezerra} V. B. Bezerra, G. L. Klimchitskaya, and V. M.
Mostepanenko, Phys. Rev. A {\bf
62}, 014102 (2000).

\bibitem{Bordag PRL} M. Bordag, B. Geyer, G. L. Klimchitskaya, and V. M.
Mostepanenko, Phys. Rev.
Lett. {\bf 85}, 503 (2000).

\bibitem{Genet}  C. Genet, A. Lambrecht, and S. Reynaud, Phys. Rev. A {\bf
62}, 012110 (2000).

\bibitem{Lambrecht} A. Lambrecht and S. Reynaud, Eur. Phys. J. D {\bf 8},
309 (2000).

\bibitem{Lamoreaux Comment} S. K. Lamoreaux, Phys. Rev. Lett. {\bf 83}, 3340 (1999).

\bibitem{Mohideen and Roy Comment} U. Mohideen and A. Roy, Phys. Rev. Lett. {\bf 83}, 3341
(1999).


\bibitem{Bordag 1998} M.~Bordag, B.~Geyer, G.~L.~Klimchitskaya,
and V.~M. Mos\-te\-panenko, { Phys. Rev.} D {\bf 58}, 075003  (1998).

\bibitem{Bordag 1999} M.~Bordag, B.~Geyer, G.~L.~Klimchitskaya,  and V.~M.
Mos\-te\-panenko,
{ Phys. Rev.} D {\bf 60}, 055004  (1999).

\bibitem{Bordag 2000} M.~Bordag, B.~Geyer, G.~L.~Klimchitskaya,   and V.~M.
Mos\-te\-panenko,
{ Phys. Rev.} D {\bf 62}, 011701(R)  (2000).


\bibitem{Most and Novello 2001} V. M. Mostepanenko and M. Novello,
Phys. Rev. D {\bf 63} 115003 (2001).

\bibitem{Most 1988} V. M. Mostepanenko and I. Yu. Sokolov, Phys. Lett. A
{\bf 132}, 313 (1988).

\bibitem{Most 1993}  V.~M.~Mostepanenko and I.~Yu.~Sokolov, Phys. Rev. D
{\bf 47}, 2882
(1993).

\bibitem{Bordag 1994} M.~Bordag, V.~M.~Mostepanenko, and I.~Yu.~Sokolov,
Phys. Lett. A {\bf 187},
35 (1994).

\bibitem{Israel} Y. N. Israelachvili and D. Tabor, Proc. R. Soc. A  {\bf 331}, 19 (1972)

\bibitem{Derjaguin}
B.~V.~Derjaguin, I.~I.~Abrikosova,
and E.~M.~Lifshitz, Quart. Rev. Chem. Soc.
{\bf 10}, 295 (1956).

\bibitem{Krause} D.~E.~Krause and E.~Fischbach, in {\it Gyros, Clocks, and
Interferometers: Testing Relativistic Gravity in Space}, edited by
C.~L\"{a}mmerzahl, C.W.F.~Everitt, F.W.~Hehl (Springer-Verlag, Berlin, 2001), pp.~292--309.

\bibitem{Blocki}  J. Blocki, J. Randrup, W. J. S\'{w}atecki, and C. F. Tsang,  Ann. Phys.
(N.Y.) {\bf 105}, 427 (1977).

\bibitem{Schaden} M. Schaden and L. Spruch, Phys. Rev. Lett. {\bf 84}, 459 (2000).

\bibitem{Bostrom} M. B\"{o}strom and B. E. Sernelius, Phys. Rev. A {\bf 61} 046101 (2000).


\bibitem{Lamoreaux 1999} S. K. Lamoreaux, Phys. Rev. A {\bf 59}, R3149 (1999).


\bibitem{Abbott} B. Abbott, {\em et al.}, Phys. Rev. Lett. {\bf 86}, 1156 (2001).

\bibitem{LEP} ALEPH Collaboration, P. Abreu, {\em et al.}, Phys. Lett. B {\bf 485}, 45
(2000); Eur. Phys. J. C {\bf 17}, 53 (2000); DELPHI Collaboration, M. Acciarri, {\em et
al.}, Phys. Lett. B {\bf 464}, 135 (1999); {\bf 470}, 268 (1999); {\bf 470}, 281 (1999); 
OPAL Collaboration, G. Abbiendi {\em et al.}, Phys. Lett. B {\bf 465}, 303 (1999); Eur.
Phys. J. C {\bf 13}, 553 (2000); {\bf 18}, 253 (2000).

\bibitem{HERA} H1 Collaboration, C. Adloff, {\em et al.}, Phys. Lett. B {\bf 479}, 358
(2000).

\bibitem{Hanhart} C. Hanhart, D. R. Phillips, S. Reddy, and M. J. Savage, Nucl. Phys. {\bf
595}, 335 (2001).

\bibitem{Barger} V. Barger, T. Han, C. Kao, and R. -J. Zhang, Phys. Lett. B {\bf 461}, 34
(1999).

\bibitem{Cullen} S. Cullen and M. Perelstein, Phys. Rev. Lett. {\bf 83}, 268 (1999).

\bibitem{Hannestad} S. Hannestad and G. G. Raffelt, hep-ph/0103201.

\bibitem{Cassisi} S. Cassisi, V. Castellani, S. Degl'Innocenti, G. Fiorentini, and B.
Ricci, Phys. Lett. B {\bf 481}, 323 (2000).

\bibitem{Hall} L. J. Hall and D. Smith, {\em Phys. Rev. D} {\bf 60}, 085008 (1999).

\bibitem{Fairbairn} M. Fairbairn, hep-ph/0101131.

\bibitem{Hannestad2} S. Hannestad, hep-ph/0102290.



\bibitem{Lambrecht personal communication} A. Lambrecht, private
communication.


\bibitem{STEP paper} E. Fischbach, S. W. Howell, S.
Karunatillake, D. E. Krause, R. Reifenberger, M. West, to appear in  Class. Quantum Grav.

\bibitem{Howell}  We thank S. W. Howell for suggesting this design.



\end{thebibliography}
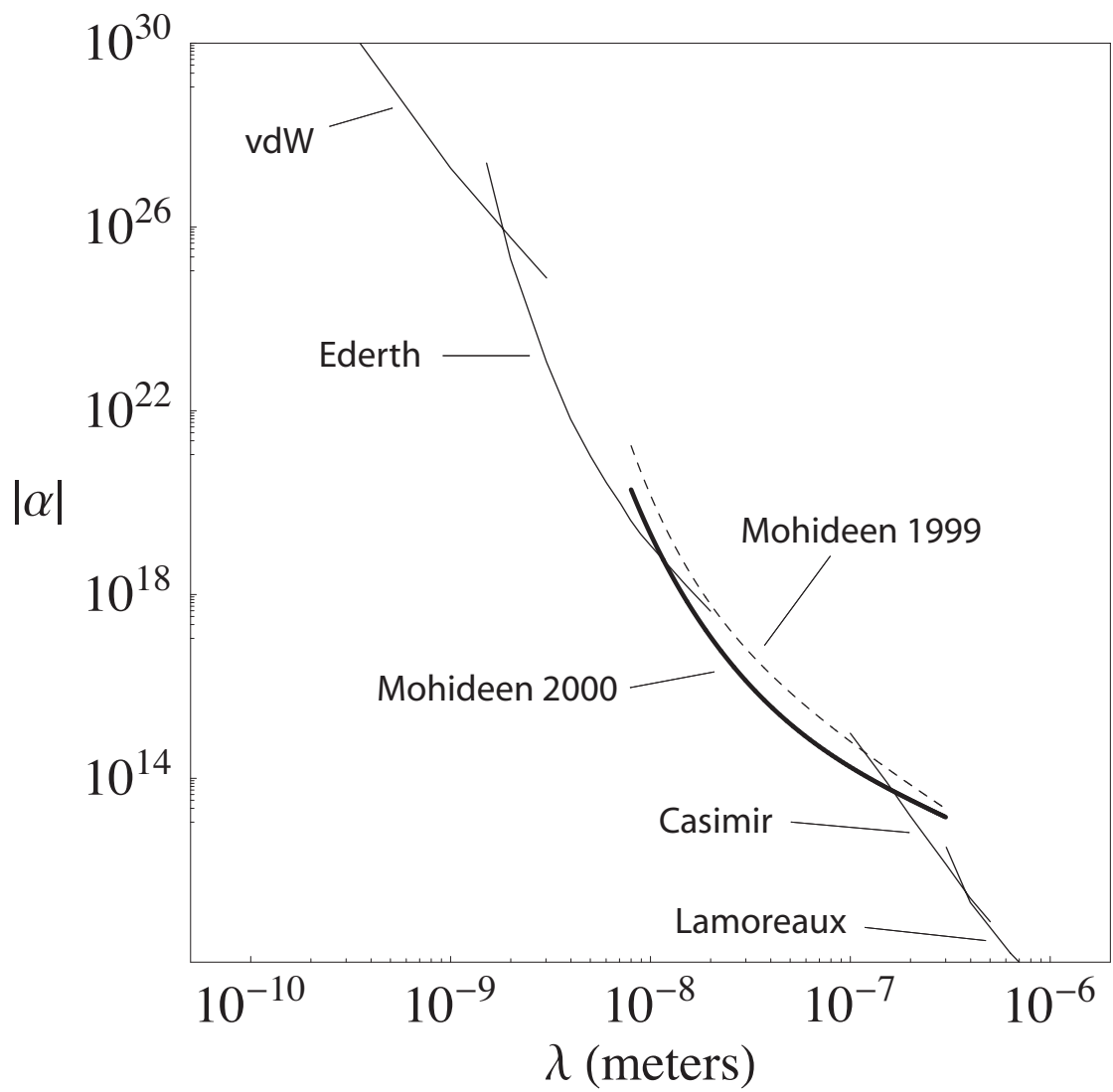
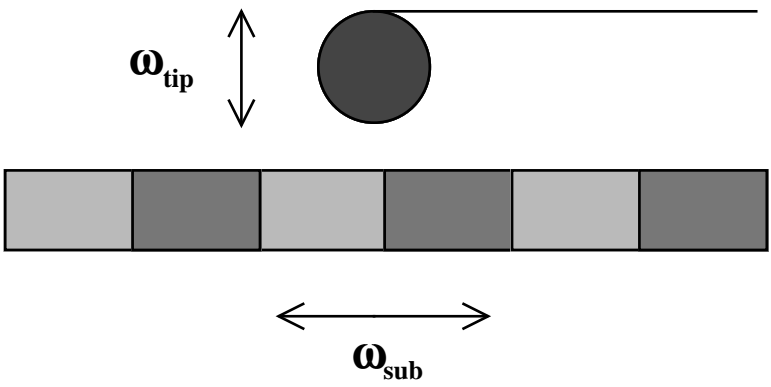
\end{document}